\documentstyle[twoside,fleqn,espcrc2,psfig]{article}

\newcommand{\AmS}{{\protect\the\textfont2
  A\kern-.1667em\lower.5ex\hbox{M}\kern-.125emS}}

\newcommand{\feh}{\hbox{$[{\rm Fe}/{\rm H}]$}}
\newcommand{\mvrr}{\hbox {${\rm M_v(RR)}$}}

\newcommand{\dv}{\hbox {$\Delta \rm V(TO-HB)$}}
\newcommand{\ea}{{\it et al.}}

\title{The Age of the Universe
\hspace*{8.8cm}\raisebox{25pt}[0pt]{\normalsize CITA-96-9} }

\author{Brian Chaboyer\address{Canadian Institute for Theoretical
Astrophysics, 60 St.\ George Street, Toronto, ON, Canada  M5S 3H8}}
\begin{document}

\begin{abstract}
Globular clusters are the oldest objects in the Galaxy whose age
may be accurately determined.  As such globular cluster ages provide
the best estimate for the age of the universe.  The age of a globular
cluster is determined by a comparison between theoretical stellar
evolution models and observational data.  Current uncertainties in the
stellar models and age dating process are discussed in detail. The
best estimate for the absolute age of the globular clusters is
$14.6\pm 1.7\,$Gyr.  The one-sided, 95\% confidence limit on the lower
age of the universe is $12.2\,$Gyr.
\end{abstract}

\maketitle

\section{Introduction}
A minimum age for the universe may be determined by obtaining a
reliable estimate for the age of the oldest objects within the
universe.  Thus, in order to estimate the age of the universe ($t_o$),
the oldest objects must be identified and dated.  The oldest objects
in the universe should contain very little (if any) heavy elements as
nucleosynthesis during the Big Bang only produced hydrogen, helium,
and lithium.  All elements heavier than lithium are not primordial in
origin, and their presence indicate that an object was not the first
to have formed in the universe.  Unfortunately, astronomers have been
unable to locate any object which consists solely of primordial
elements.  There are, however, objects which contain small amounts of
the heavier elements.  These objects will be the focus of my talk.

Within our galaxy, the oldest objects which can be dated are the
globular clusters (GCs).  GCs are compact stellar systems, containing
$\sim 10^5$ stars (see figure 1).  These stars contain few heavy
elements (typically 1/10 to 1/100 the ratio found in the Sun).  GCs
are spherically distributed about the Galactic center, suggesting that
GCs were formed soon after the proto-Galactic gas started collapsing.
Thus, GCs were among the first objects formed in the Galaxy.  An
estimate of their age will provide a reasonable lower limit to the age
of the universe.  In order to estimate the true age of the universe,
one must add to the age of the GCs the time it took GCs to form after
the Big Bang.
\newpage

\vspace*{5pt}
\noindent
\centerline{\psfig{figure=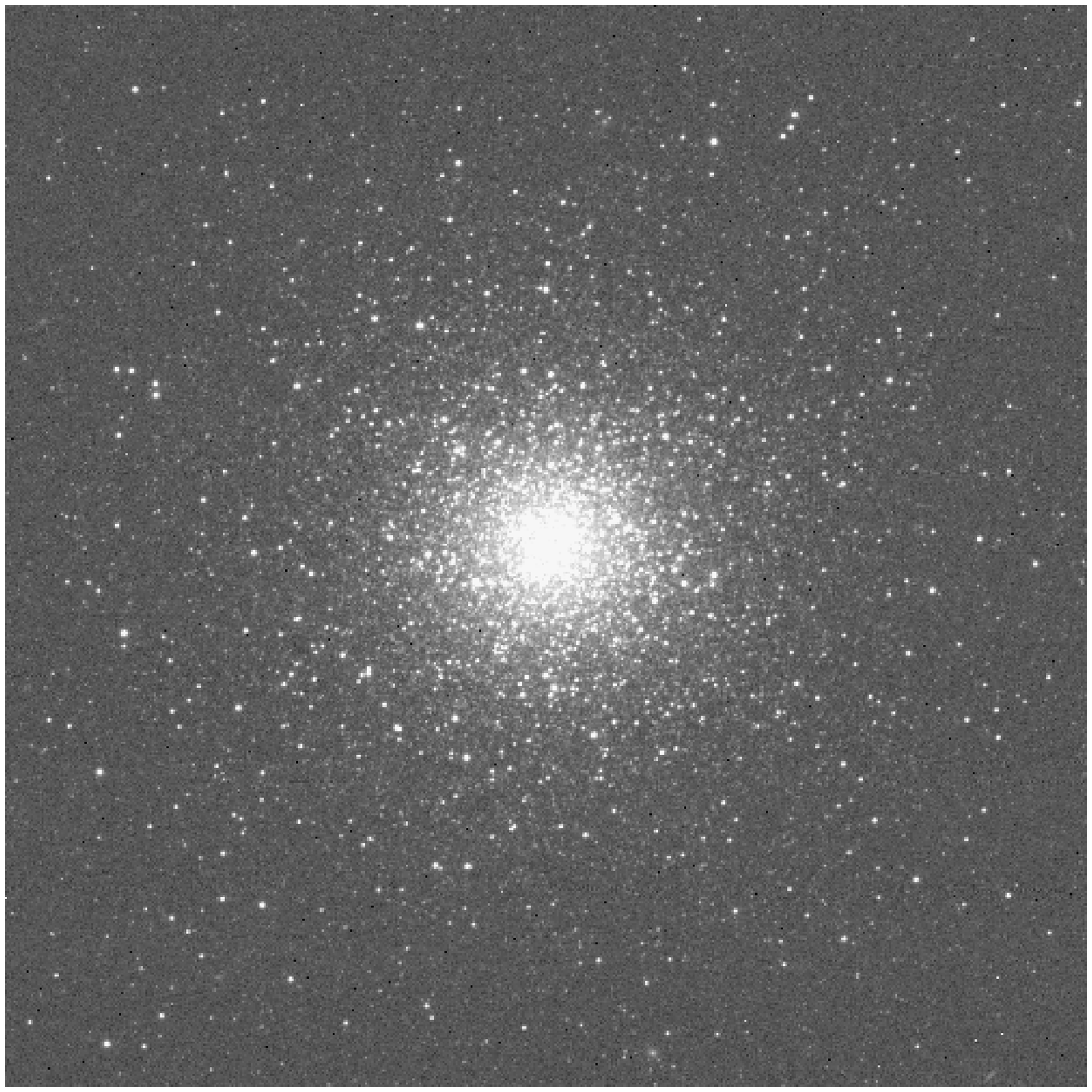,height=7.5cm}}
\vspace*{5pt}

\noindent
Figure 1.\ An example of a typical GC, M53. This cluster
contains roughly $10^6$ stars.  Image courtesy of Ata Sarajedini.
\vspace*{10pt}

\noindent

It is important to realize that the estimate for $t_o$ is based on a
single method, applying stellar evolution models to GC observations.
This is in sharp contrast to estimates for the Hubble constant
($H_o$), which are based on a wide variety of independent techniques
(see the review by Trimble in this volume).  Whereas estimates for the
Hubble constant may differ by a factor of two, depending on the
observer and the technique which is used, estimates for the absolute
age of the GCs typically agree with each to within $\sim 10\%$.  The
important considerations then, are to estimate the uncertainty in the
stellar models and the derived value of $t_o$, and to test the stellar
models in as many ways as possible to ensure that no systematic errors
exist.

The study of $t_o$ has taken on increasing significance in recent
years, as the value of $t_o$ derived from GCs appears to be longer
than the expansion age of the universe, derived from $H_o$.  The value
of $H_o t_o$ is a function of the the cosmological constant
($\Lambda$) and density of the universe ($\Omega$, in units of the
critical density).  For a `standard' inflationary universe ($\Omega =
1$, $\Lambda = 0$), $H_o t_o = 2/3$.  A value of $H_o t_o > 1$
requires a non-zero cosmological constant, or a significant revision to
standard Big Bang cosmology.

In this review I will describe in detail how GC ages are estimated.
Some basic observational properties of GCs are summarized in \S
\ref{sec2}.  The construction of stellar models which are used to date
GCs are described in \S \ref{sec3}, while \S \ref{sec3a} contains a
discussion of age determination techniques for GCs.  Section
\ref{sec4} contains a detailed discussion of possible errors in the
age estimates for GCs.  This section includes the results of a recent
Monte Carlo analysis which has resulted in a firm lower limit to the
age of the universe.  Various tests of stellar models are presented in
\S \ref{sec5}, including white dwarf cooling time-scales.  Finally, \S
\ref{sec6} contains a summary of this review.


\section{Globular Cluster Observations}\label{sec2}
Observers typically measure the apparent magnitude ($\propto
2.5\,\log(luminosity)$) of as many stars as possible within a
GC. These measurements are usually taken through at least two
different filters, so that the apparent colour of the
stars may also be determined.  The fact that a GC contains a large
number of stars all at the same distance from the Earth is an enormous
advantage in interpreting the observations.  The ranking in apparent
luminosity (as seen in the sky) is identical to the ranking by
absolute luminosity.  Figure 2 is a typical example of how the
observations are reported, as a colour-magnitude  diagram.
\newpage
\vspace*{5pt}

\noindent
\vspace*{1pt}
\centerline{\psfig{figure=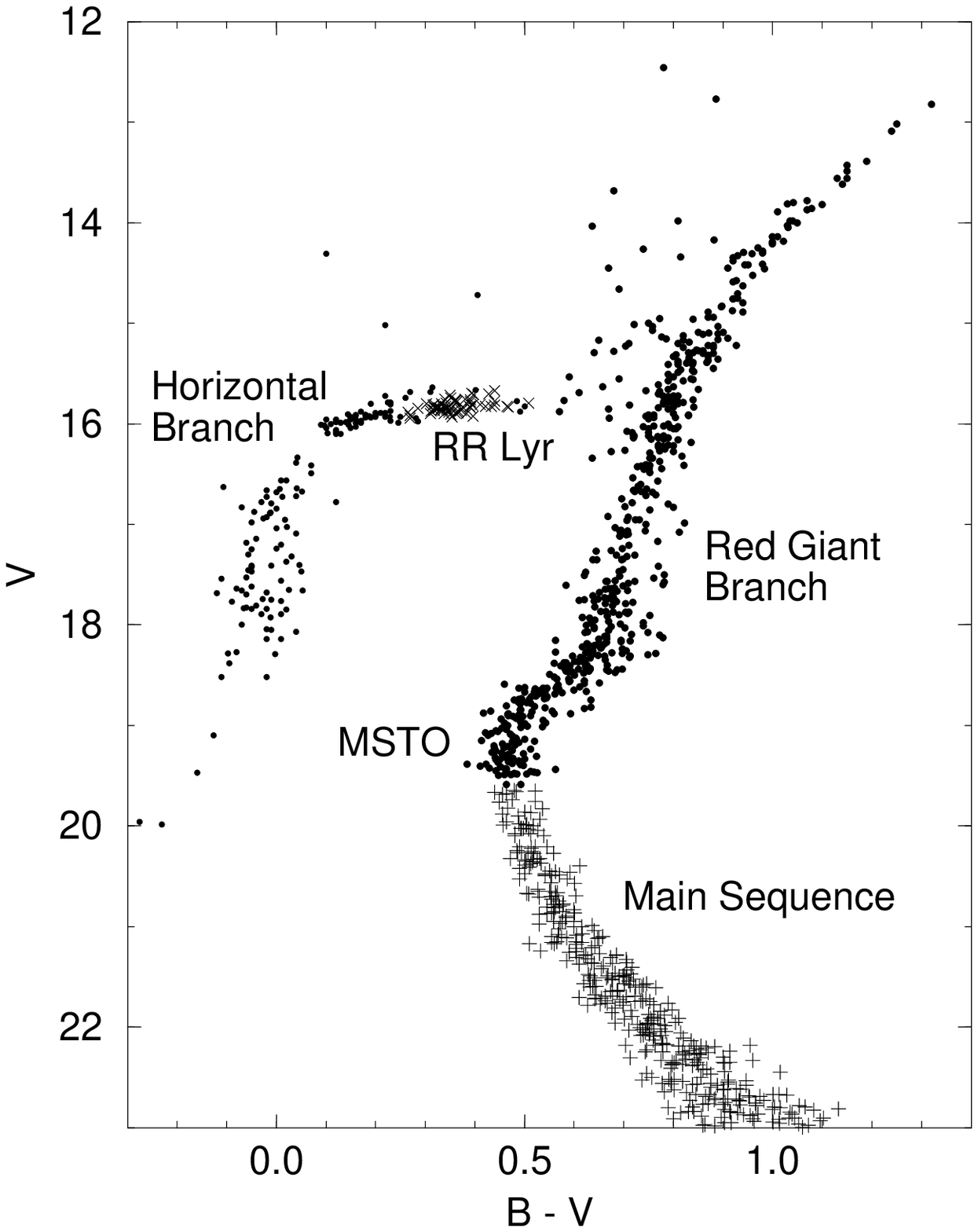,height=8.5cm}}
\vspace*{-1pt}

\noindent
Figure 2.\ A colour-magnitude diagram of a typical GC, M15 [1].
The vertical axis plots the magnitude (luminosity) of the star in the
V wavelength region, with brighter stars having smaller magnitudes.
The horizontal axis plots the colour (surface temperature) of the
stars, with cooler stars towards the right.  The various evolutionary
sequence have been labeled (see text).  For clarity, only about 10\%
of the stars on the main sequence have been plotted.
\vspace*{17pt}

\noindent

One of the great triumphs of the theory of stellar structure and
evolution has been an explanation of the colour-magnitude diagram.
The more  massive a star, the quicker it burns its nuclear fuel and
evolves.  Thus, stars of different initial masses will be in different
stages of evolution. Figure 2 graphically illustrates the various
phases of stellar evolution.  After their birth, stars start
to burn hydrogen in their core.  This is referred to 
as the {\it main sequence\/}.  A star will spend approximately 90\% of
its life on the main sequence. The Sun is a typical example of a main
sequence star.  Eventually, a star exhausts the supply of hydrogen in
its core, and reaches the {\it main sequence turn-off \/} (MSTO).
This point is critical in the age determination process.  After a star
has burned all of the hydrogen in its core, the outer layers 
expand and hydrogen fusion occurs in a shell surrounding the helium
core.  The expansion of the outer layers causes the star to cool and
become red, so stars in this phase of evolution are said to occupy the
{\it red giant branch\/}.  The hydrogen burning shell moves out in
mass coordinates, leading to increasing luminosity and helium core
mass.  On the red giant branch, a typical GC star is believed to lose
$\approx 25\%$ of its mass.  When and how this occurs is a still a
subject of research.  Eventually, the helium core becomes so dense
that helium fusion is ignited.  The star quickly settles onto the {\it
horizontal branch\/} (HB).  On the HB, fusion of helium occurs in the
core, surrounded by a shell of hydrogen fusion.  Exactly where a star
lies on the horizontal branch (blue or red) depends on how much mass
loss has occurred on the red giant branch.  Some stars on the
horizontal branch are unstable to radial pulsations --- these stars
are referred to as {\it RR Lyrae\/} stars.  A star's lifetime on the
HB is extremely short, it soon exhausts the supply of helium at its
core and becomes an asymptotic giant branch star (similar to the red
giant branch), burning helium and hydrogen in shells about a carbon
core.  In GC stars (like the Sun), the core temperatures and densities
never become high enough to ignite the fusion of carbon.  After a star
finishes its helium and hydrogen shell burning, the envelope may be
ejected, while the core contracts and becomes extremely dense. The
star becomes dim, as the only energy available to the star is that
from gravitational contraction. In this terminal phase of evolution,
the star is referred to as a {\it white dwarf\/} (not shown in Fig.\
2).  When the Sun becomes a white dwarf, its radius will be similar to
the Earth's.  As a white dwarf ages, it continues to cool, and emit
less radiation.  Ultimately, a star will reach equilibrium with its
surroundings, becoming virtually invisible.

\section{Stellar Models}\label{sec3}
Our understanding of stellar evolution is based on stellar structure
theory.  There are numerous textbooks which describe the basic theory
of stellar structure and the construction of stellar models (e.g.\
\cite{schw}).  A stellar model is constructed by solving the four
basic equations of stellar structure: (1) conservation of mass; (2)
conservation of energy; (3) hydrostatic equilibrium and (4) energy
transport via radiation, convection and/or conduction.  These four,
coupled differential equations represent a two point boundary value
problem.  Two of the boundary conditions are specified at the center
of the star (mass and luminosity are zero), and two at the surface.
In order to solve these equations, supplementary information is
required.  The surface boundary conditions (temperature and pressure)
are based on stellar atmosphere calculations.  The equation of state,
opacities and nuclear reaction rates must be known.  The mass and
initial composition of the star need to be specified.  Finally, as
convection can be important in a star, one must have a theory of
convection which determines when a region of a star is unstable to
convective motions, and if so, the efficiency of the resulting heat
transport.  Once all of the above information has been determined a
stellar model may be constructed by solving the four stellar structure
equations.  The evolution of a star may be followed by computing a
static stellar structure model, updating the composition profile to
reflect the changes due to nuclear reactions and/or mixing due to
convection, and then re-computing the stellar structure model.

There are a number of uncertainties associated with stellar evolution
models, and hence, age estimates based on the models.  Probably the
least understood aspect of stellar modeling is the treatment of
convection.  The understanding of convection in a compressible plasma,
where significant amounts of energy can be carried by radiation, is a
long standing problem.  Numerical simulations hold promise for the
future \cite{kim}, but at present one must view properties of stellar
models which depend on the treatment of convection to be uncertain,
and subject to possibility large systematic errors. Main sequence, and
red giant branch GC stars have surface convection zones. Hence, the
surface properties of the stellar models (such as its effective
temperature, or colour) are rather uncertain.  Horizontal branch stars
have convective cores, so the predicted luminosities and lifetimes of
these stars are subject to possible systematic errors.

Another important consideration in assessing the reliability of
stellar models, and the ages they predict for GCs is that the advanced
evolutionary stages are considerably more complicated than the main
sequence.  Thus, one may expect that the main sequence models are
least likely to be in error.  Observations of CNO abundances in red
giant branch stars indicate that some form of deep mixing occurs in
these stars, which is not present in the models \cite{langer}. In
contrast, there is no observational evidence suggesting that the low
mass, main sequence models are in serious error.  For this reason, age
indicators which are based on main sequence models are the most
reliable.

\section{Globular Cluster Age Estimates}\label{sec3a}
A theoretical stellar evolution model follows the time evolution of a
star of a given initial composition and mass.  Stars in a given GC all
have the same chemical composition and age (with a few exceptions),
but different masses.  Thus, in order to determine the age of a GC, a
series of stellar evolution models with the same composition but
different masses must be constructed.  Interpolation among these
models yields an {\it isochrone}, a theoretical locus of points for
stars with different masses but the same age.  The theoretical
calculations are performed in terms of total luminosity, and effective
temperature.  As discusses in \S \ref{sec2}, observers measure the
brightness of a star over specified wavelength ranges.  Thus, it is
necessary to convert from the effective temperature and total
luminosity to luminosities in a few specified wavelength
intervals. This requires a detailed knowledge of the predicted flux,
as a function of wavelength.  Theoretical stellar atmosphere models
are used to perform this conversion.  The result of such calculations
is illustrated in Figure 3, which plots isochrones for 3
different ages, in terms of absolute V magnitude, and B--V colour.
\begin{figure}[t] \label{fig3}
\vspace*{20pt}
\centerline{\psfig{figure=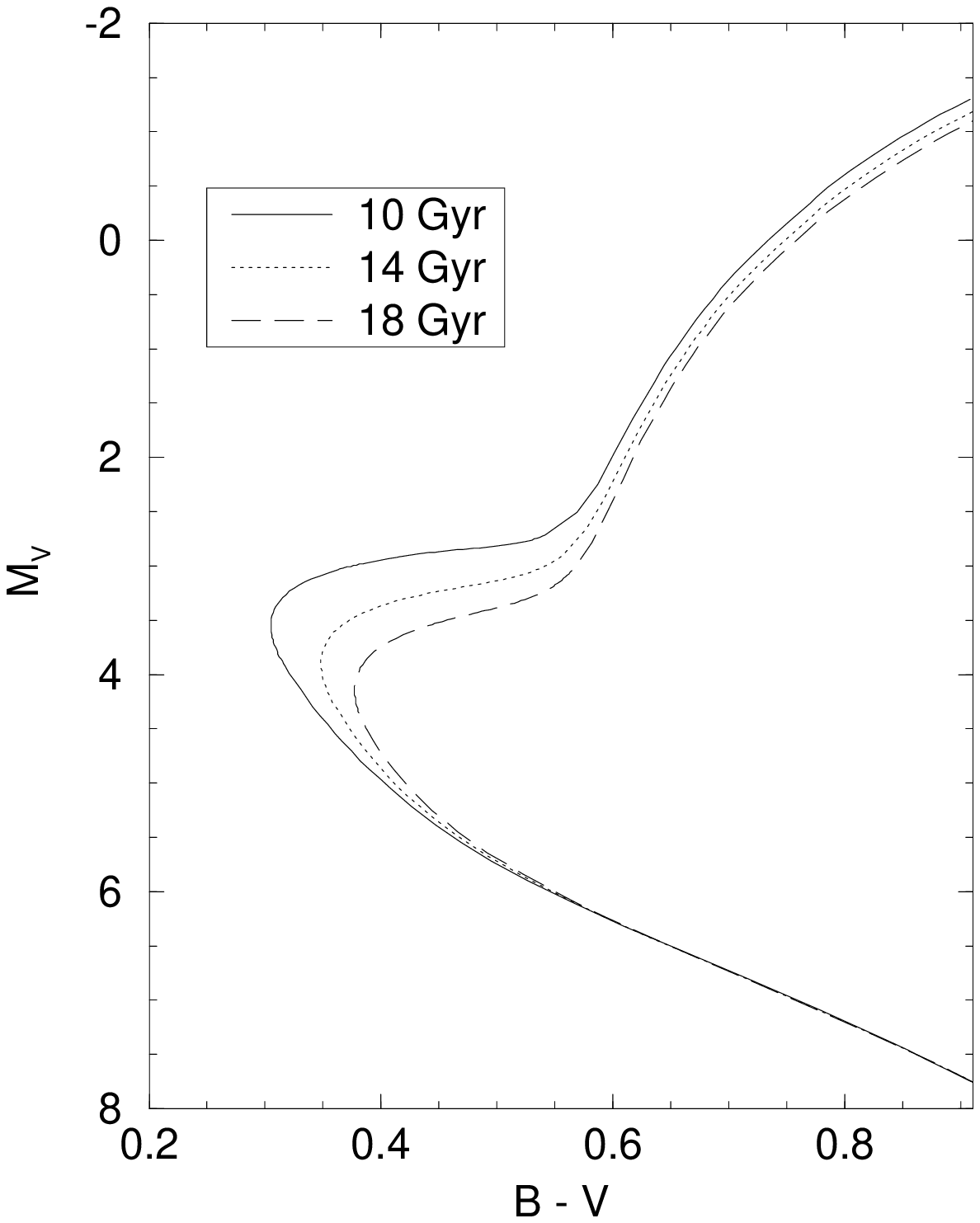,height=9.0cm}}
\vspace*{-20pt}

\caption{Theoretical isochrones for 3 ages (10, 14 \& 18 Gyr).  Note
that the main sequence turn-off becomes fainter and redder for older
isochrones. }
\end{figure}
As can be seen in Figure 3, differences in age lead to large
differences in the MSTO region.  The MSTO becomes fainter, and redder 
as the age increases.  

In order to provide the tightest possible constraints on the age of
the universe, it is important to use an age indicator which has the
smallest possible theoretical error.  From the discussion in \S
\ref{sec3}, it is clear that (a) the main sequence is the best
understood phase of stellar evolution and (b) the predicted
luminosities of the models are better known than the colours.  These
two reasons, coupled with the age variation exhibited in Figure
3 lead to the conclusion that the absolute magnitude of the
MSTO is the best indicator of the absolute age of GCs \cite{renzini}.
Unfortunately determining the magnitude of the MSTO in observational
data is quite difficult, as the MSTO region extends over a large range
in magnitude (see Fig.\ 2).  For this reason, it is best to determine
the mean age of a large number of GCs, in order to minimize the
observational error.

Observers measure the apparent magnitude of a star. In order to
convert to the absolute magnitude (and hence, determine an age), the
distance to a GC must be determined.  Obtaining the distance to an
object remains one of the most difficult aspects of astronomy.  At
present, there are two main techniques which are used to determine the
distance to a GC (1) main sequence fitting to local sub-dwarfs with
well measured parallaxes, and (2) using the observed magnitude of the
HB combined with a relationship for the absolute magnitude of the HB
(derived using RR Lyrae stars).  Unfortunately, there are few
sub-dwarfs with well measured parallaxes (a situation which should
change once date from the Hipparcos satellite are released), so at the
present time the use of the HB to set the distance scale to GCs is the
most reliable.  The HB has the advantage that the difference in
magnitude between the MSTO and the HB (\dv) is independent of
reddening.  Thus, \dv\ is a widely used age determination technique,
which uses the absolute magnitude of the main sequence turn-off as its
age diagnostic.  There are a number of observational and theoretical
techniques which may be used to obtain the absolute magnitude of the
RR Lyr stars (\mvrr), with the general conclusion that $\mvrr = \mu\,
\feh + \gamma$ where $\mu$ is the slope with metallicity and $\gamma$
is the zero-point \cite{carney}.  Uncertainties in the slope primarily
effect the relative ages of GCs.  When a number of GCs are studied,
the uncertainty in the slope has a negligible effect on the derived
mean age.  The uncertainty in the \mvrr\ zero-point has a large impact
on the derived ages (see \S \ref{sec4}).

A number of different researchers have constructed stellar
evolutionary models and isochrones which they have used to estimate
the age of the GCs.  These estimates agree well with each other, and
indeed have remained relatively constant since $\sim 1970$
\cite{ageref}.  This is not too surprising, as the basic assumptions
and physics used to construct the stellar models are the same for the
different research groups, and have not changed for a number of years.
These studies have also revealed that GCs are not all the same age.
Thus, to provide the best estimate for the age of the universe one
must select a sample of the oldest GCs.  Chaboyer, Demarque, Kernan \&
Krauss (hereafter CDKK, \cite{cdkk}) have recently completed a study
of the absolute age of the oldest GCs, which they found to be
$14.6\,$Gyr.  The novel aspect of this work was the detailed
consideration of the possible sources of error in the stellar models
and age determination process which allowed CDKK to provide an estimate
of the error associated with their age estimate.

\section{Error Estimates}\label{sec4}
To access the error in the absolute GC age estimates, one must review
the assumptions and physics which are used to construct stellar models
and isochrones. The discussion presented in \S \ref{sec3} and \S
\ref{sec3a} of the GC ages determination process allows one to compile
a list of possible sources of error in theoretical calibration of age
as a function of the absolute magnitude of the MSTO:
\begin{enumerate}
\item assumption of hydrostatic equilibrium in radiative regions
\item nuclear reaction rates
\item opacities
\item equation of state
\item treatment of convection
\item surface boundary conditions
\item chemical composition
\item conversion from theoretical luminosities to observed magnitudes
\end{enumerate}
This is a lengthly list, which has been studied in some detail
\cite{cdkk,myage}.   In this review, I will concentrate on a few items
which turn out to be particularly important, or for which improved
calculations have recently become available.  

The validity of the assumption of hydrostatic equilibrium in the
radiative regions of stars has received considerable attention. Not
surprisingly, if some process operates which mixes material into the
core of a star, the main sequence life-times will be prolonged, and
hence the true age of the GCs will be older than current estimates.
However, if a microscopic diffusion is active (causing helium to sink
relative to hydrogen), than the main sequence life-times are
shortened, leading to lower estimates for the age of the GCs
\cite{vand}.  The inclusion of diffusion has been found to lower the
GC age estimates by 7\%.  There is evidence from helioseismology that
diffusion is occurring in the sun \cite{christian}, but models of halo
stars which incorporate diffusion are unable to match the Li
observations in halo stars \cite{chabli}.  Until this matter is
resolved, GC age estimates are subject to a possible 7\% systematic
error due to the effects of diffusion.

The correctness of the nuclear reaction rates used in stellar models
have been extensively analyzed due to their importance in the solar
neutrino problem \cite{bahcall}.  Although predicted solar neutrino
fluxes are quite sensitive to possible errors in the nuclear reaction
rates, they have a minor effect on GC age estimates \cite{myage}.

There has been considerable effort devoted to determining the
opacities used in stellar models.  A number of different research
groups have calculated opacities, using independent methods.  The
agreement between these calculations, particularly for metal-poor
mixtures is quite good \cite{opac}.  Indeed, the high-temperature
opacities in metal-poor stars have only changed by $\sim 1\%$ between
the mid-1970's and the present.  

The equation of state used in stellar models is another area of active
research.  Detailed calculations have lead to the availability of
equation of state tables which are a considerable improvement over the
simple analytical formulae usually employed in stellar evolution
calculations \cite{rogers}.  These calculations include Coloumb
effects, which had typically been ignored in previous calculations.
It has been found that the improved equation of state reduces GC age
estimates by 7\% \cite{chabeos}. Independent calculations (using
entirely different physical assumptions) \cite{opacity} lead to an
equation of state which agrees quite well with \cite{rogers}.  Thus,
it is unlikely that there are significant errors in the new equation
of state calculations.

The correct composition to use in stellar models is a long standing
problem.  The helium mass fraction is taken to be the primordial
value \cite{KK1}, $Y = 0.23 - 0.24$.  A generous $\pm 0.03$
uncertainty in $Y$ leads to a negligible uncertainty in the derived age
\cite{myage}.  In contrast, the uncertainty in the heavy element
composition leads to a significant error in the derived age.  It is
relatively easy to determine the abundance of iron, and it was
generally assumed that the other heavy elements present in a GC
star would be scale in a similar manner.  However, 
it has become clear that the $\alpha$-capture elements (O, Mg, Si, S,
and Ca) are enhanced in GC stars, relative to the ratio
found in the Sun \cite{lambert}.  Oxygen is the most important of
these elements, being by far the most abundant.  However, it is quite
difficult to determine the abundance of oxygen in GC
stars \cite{oxygen}, and as a consequence the uncertainty in the
oxygen abundance is leads to a relatively large error ($\pm 6\%$) in
the derived age of the oldest GCs \cite{myage}.

In an attempt to take into account all of the possible uncertainties in
the GC age dating process, CDKK have made a detailed examination of
the likely error in each of the parameters.  CDKK performed a
Monte-Carlo simulation, in which the various quantities were allowed
to vary within some specified distribution, chosen to encompass the
possible uncertainty in that quantity.  The mean age of
the 17 oldest GCs was determined using 1000 independent sets
of isochrones.  Assuming that the distances to the GCs are known
exactly, a mean age of $14.6\pm 1.1\,$Gyr was determined, with a 95\%
confidence limit that  GCs are older than $12.9\,$Gyr.  Allowing
for an uncertainty in the distance modulus ($\pm 0.08\,$mag in the RR
Lyrae zero-point) increases the allowed range to $14.6\pm 1.6\,$Gyr,
with a 95\% confidence limit that the GCs are older than $12.1\,$Gyr.
The error in the distance modulus dominates the overall
uncertainty in the absolute age of the oldest GCs.

\section{Tests of Stellar Models}\label{sec5}
The discussion presented in the previous section, and the error
analysis performed by CDKK assumes that there are no unknown
systematic errors in the GC age determination process.  This
assumption is supported by four independent tests of stellar structure
theory:
\begin{enumerate}
\item fitting theoretical isochrones to observed GC
colour-magnitude diagrams;
\item comparison between observed and predicted luminosity functions;
\item observations of solar $p$-modes which probe the structure of the
Sun down to $r = 0.05\,{\rm R}_\odot$; and 
\item white dwarf age estimates for the local Galactic disk which
agree with MSTO age estimates.
\end{enumerate}
These tests are summarized below.

Theoretical isochrones provide a good match to observed GC
colour-magnitude diagrams (Figure 4).  The freedom to
modify the predicted colours of the models (due to our poor treatment
of the surface convection in these stars), implies that this is not a
definitive test of stellar models which proves that they are correct.
However, the absence of any unexplained features in the observed
colour-magnitude diagram constrains non-standard models.  For example,
models which include a mixed core (which predict older ages for
GCs) predict a `hook' in the MSTO region which is not
observed.  Hence, one may conclude that GC stars do not
have cores which have been extensively mixed.
\begin{figure}[t] \label{fig4}
\centerline{\psfig{figure=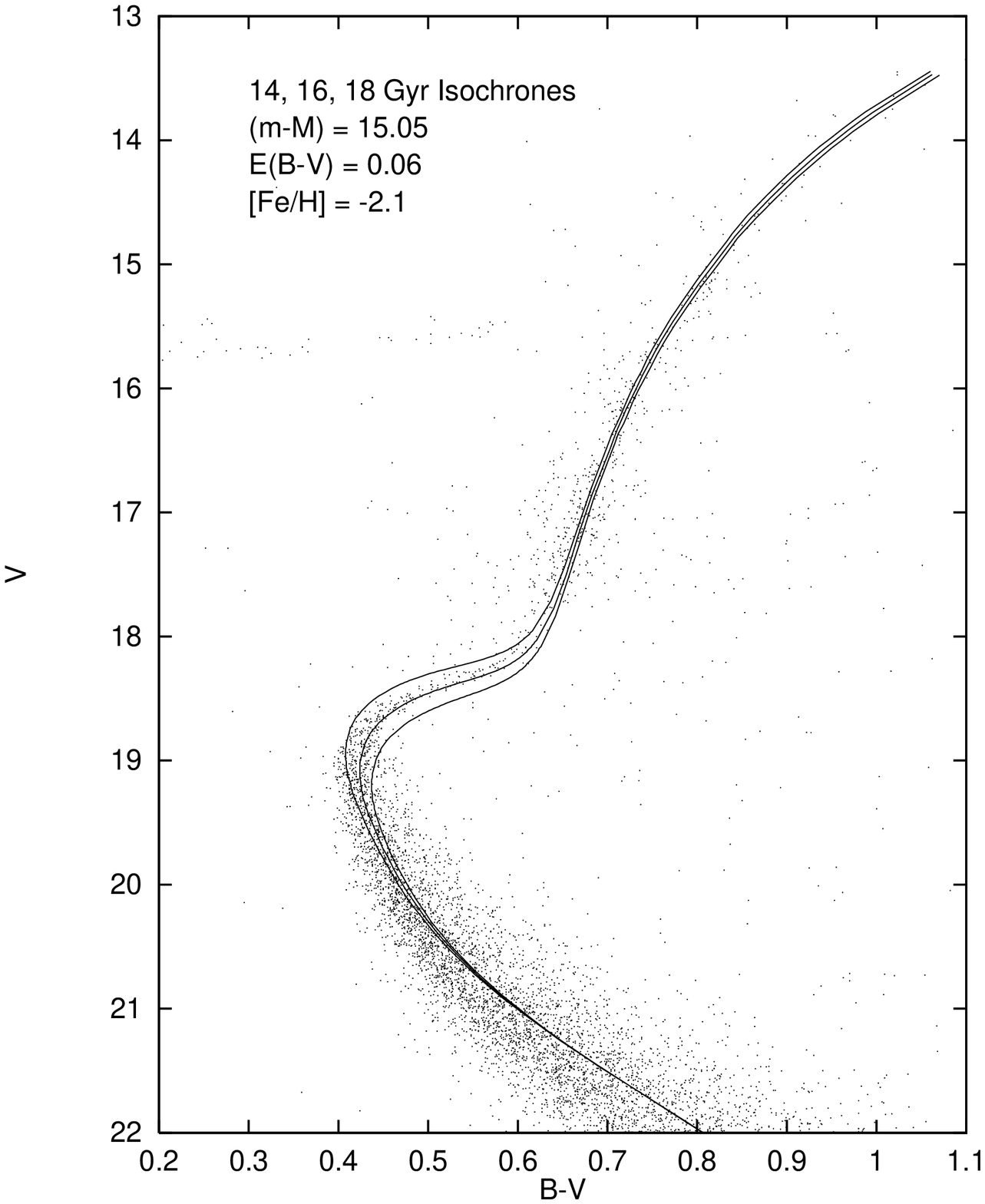,height=9.0cm}}
\vspace*{-10pt}

\caption{Theoretical isochrones with ages of 14, 16 and 18 Gyr are fit
to the GC M68 [21].  These isochrones do a good job of
matching the observed colour-magnitude diagram.}
\end{figure}

The number of stars as a function of luminosity is referred to as a
luminosity function (LF).  On the lower main sequence, the LF is a
function of the number of stars per unit mass, and the mass-luminosity
relationship.  In the more advanced evolutionary stages (starting
about 1 magnitude below the MSTO), the evolutionary time-scales are
very short, and dominate the number counts.  Hence, observed LFs
provide an excellent test of the {\em relative} lifetimes predicted by
the stellar models.  The freedom to choose an overall normalization
factor between the observations and theory implies that this is not a
test of the absolute lifetimes.  In general, a good match is found
between predicted and observed LFs \cite{lfweiss} (see Figure
5), implying that the relative evolutionary time-scales
predicted by the models are correct. Thus, any mechanism which
shortens the main sequence lifetime of GC stars (and hence, shortens
the GC age estimates) must predict a corresponding decrease in the
more advanced evolutionary phases, like the RGB.  There are
suggestions with the present data sets that models which incorporate
isothermal cores do not match the observations \cite{lfweiss},
although the conclusions are not definitive.  The isothermal core
models predict GC ages which are about 20\% smaller than the standard
stellar models \cite{fs}.  It is now technically possible to obtain
much larger observational data sets, which will lead to much smaller
(Poisson) error bars, and a more definitive test of the relative
lifetimes predicted by stellar evolution models.
\begin{figure}[t] \label{fig5}
\vspace*{22pt}
\centerline{\psfig{figure=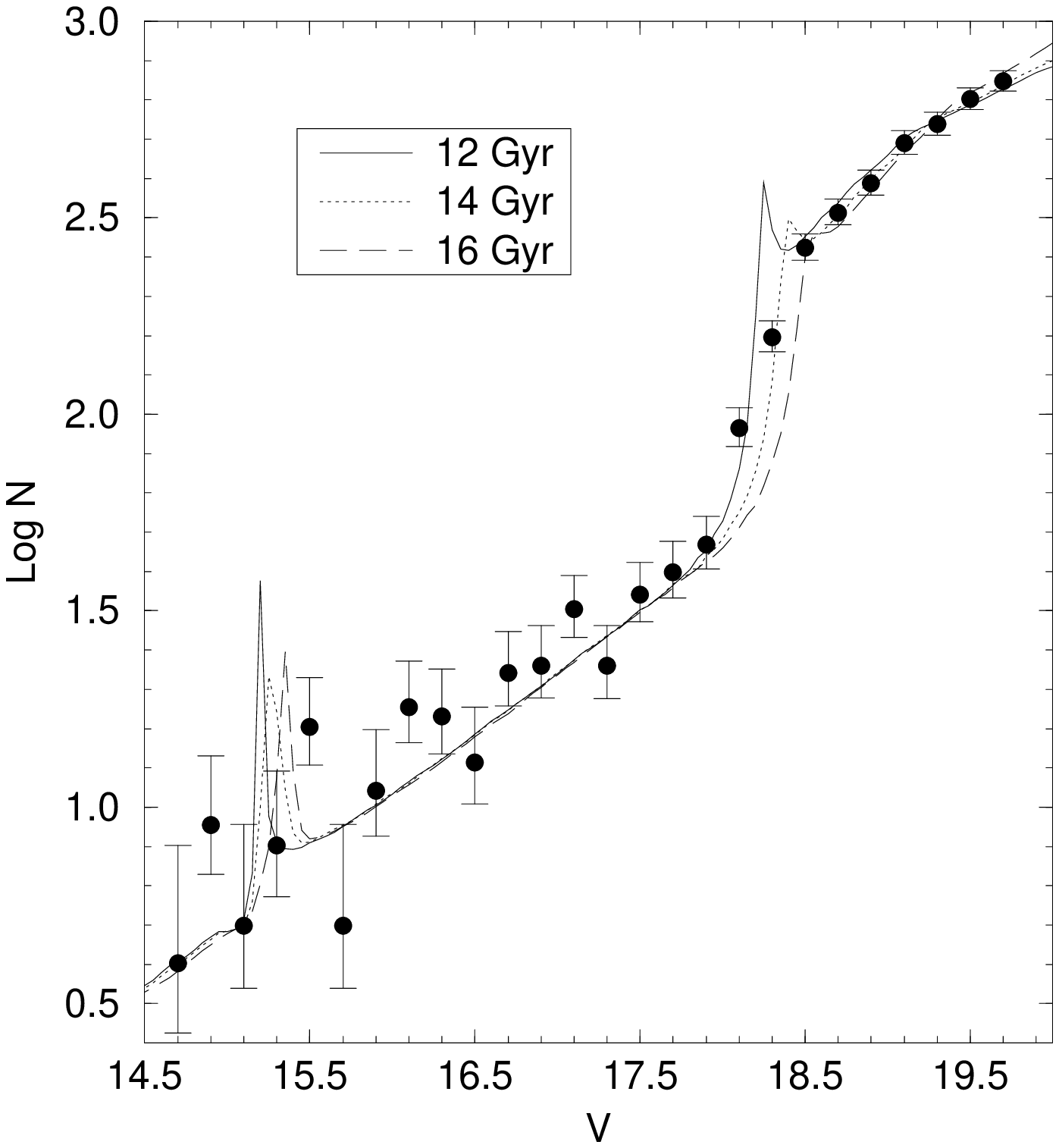,height=7.5cm}}
\vspace*{-20pt}

\caption{Theoretical luminosity functions with ages of 12, 14 and 16
Gyr are fit to the GC NGC 288 [23].  The luminosity
functions do a good job of matching the observed data.}
\end{figure}

The internal structure of the Sun is predicated to be quite similar to
a typical main sequence GC star.  Both stars fuse hydrogen via the pp
cycle in their radiative interiors and have a surface convection
zone.  Hence, tests of solar models may also be viewed as tests of the
stellar models which are used to determine GC ages.  Millions of
non-radial oscillatory modes have been observed at the surface of the
Sun.  These non-radial modes are referred to as $p$-modes, and provide
an unique test of stellar evolution. Precise observations of the
frequencies of the $p$-modes make it possible to infer many properties
of the solar interior and to test stellar evolution models
\cite{jcd}. For example, helioseismology has lead to estimates for the
solar helium abundance \cite{basu} and has put strict limits on the
amount of overshoot present below the surface convection zone
\cite{conv2}.  Inversions of the observed frequencies with respect to
a solar model yield the difference in the squared sound speed between
the model and the Sun.  The squared sound speed is proportional to the
pressure divided by the density, $c^2 \propto P/\rho$.  Thus,
inversions of the solar $p$-modes are a direct test of the interior
structure predicted by solar models.  The results of such an inversion
are shown in Figure 6.  The agreement is remarkable good, with
differences of less than 0.5\% throughout most of the model.  The
inversions do not extend to the very center of the star -- the
observed $p$-modes do not penetrate below $r \sim 0.05\,{\rm
R}_\odot$, and so do not probe the structure of the sun below this
point.  The $p$-mode observations indicate that the surface structure
of the models are in error (implying that the treatment of convection
needs to be improved).
\begin{figure}[t] \label{fig6}
\vspace*{17pt}
\centerline{\psfig{figure=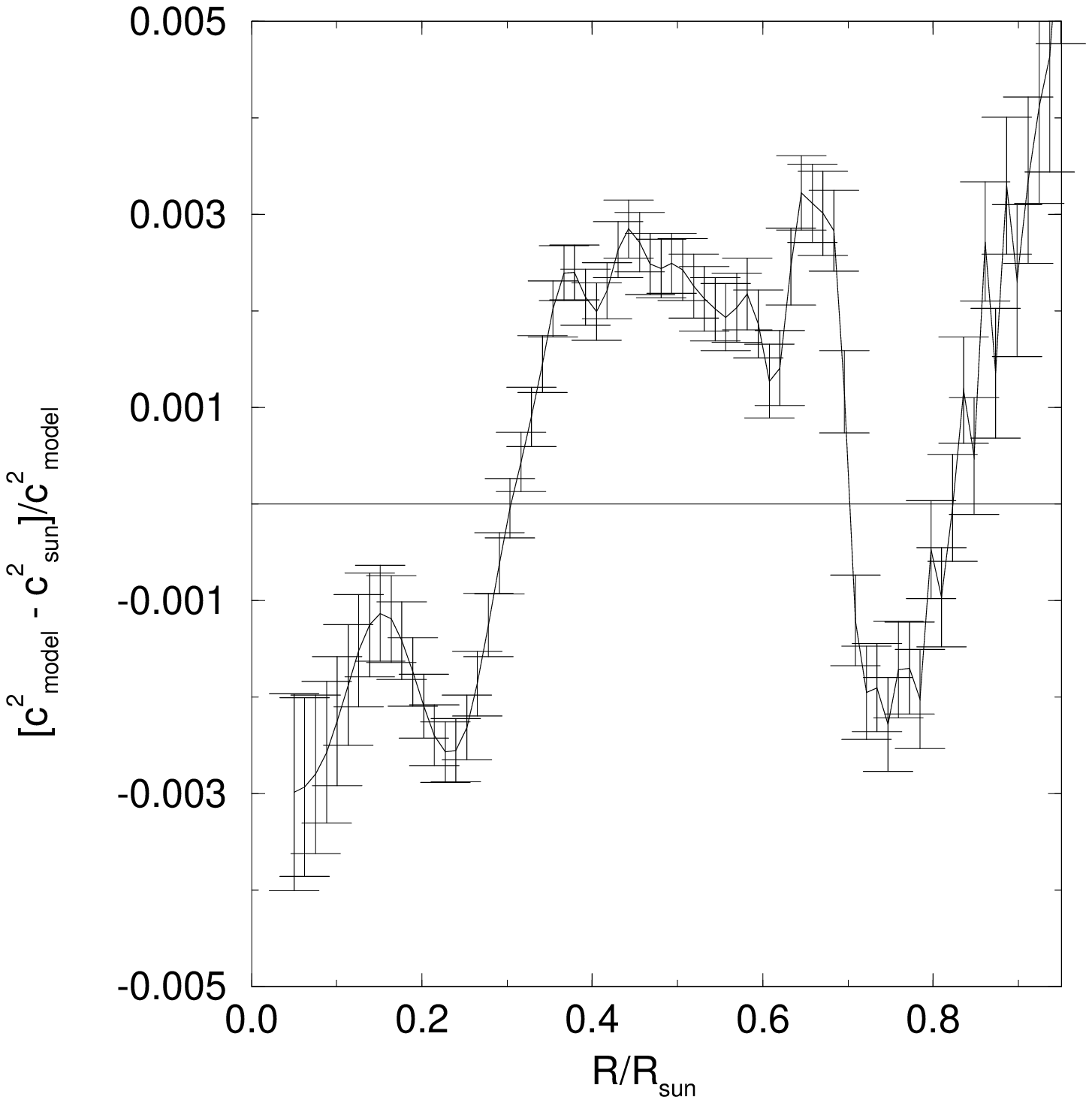,height=7.8cm}}
\vspace*{-20pt}

\caption{The difference in the square of the sound speed, between a
reference solar model and the actual Sun, as a function of radius.
This plot is obtained by inverting the observed frequencies of the
solar $p$-modes with respect to the reference model.}
\end{figure}

The excellent agreement between the sound speed in the Sun, and that
predicted by solar models is strong evidence that there are no serious
errors in current stellar evolution models.  However, there remains
the long standing discrepancy between the predicted solar neutrino
fluxes and those observed on the earth.  Four independent neutrino
experiments have observed a solar neutrino flux which is $1/2 - 1/3$
the predicted value \cite{bahcall2}.  A solution to this problems
requires either (a) new neutrino physics, or (b) a systematic error in
the stellar evolution models. Given the excellent agreement with
helioseismology, and the apparent energy dependence of the observed
solar neutrino deficient, it is likely that a resolution of the solar
neutrino problem requires new neutrino physics \cite{neut}.  However,
until definitive observational evidence is obtained \cite{sno}, there
remains a possibility that there is some unknown systematic error in
the solar models.  If this were to be the case, then our estimates for
GC ages would require revision.

Estimates for the age of a stellar population may be obtained from
white dwarf cooling curves.  The assumptions and physics used to
construct white dwarf cooling models are quite distinct from those
used in stellar evolution models.  Hence, white dwarf cooling curves
provide an independent test of the lifetimes predicted by stellar
evolution models.  The basic idea behind white dwarf cooling curve age
estimates is that as a white dwarf ages, it becomes fainter, and
cooler.  Thus, for a given age white dwarfs will not exist below a
minimum temperature and luminosity.  White dwarf cooling curves are
relatively simple to model, but it is difficult to observe white
dwarfs due to their low luminosity.  Indeed, at the present time it is
impossible to detect the faint end of the white dwarf luminosity
function in GCs.  The turn-over in the white dwarf luminosity function
has been detected in the local solar neighborhood, and provides an
independent estimate for the lifetime of the Galactic disk of
$10.5^{+2.5}_{-1.5}\,$Gyr \cite{wood}. This is in agreement with
estimates for the age of the oldest open clusters in the disk, 7 --- 9
Gyr \cite{open} which are based on MSTO ages.  This suggests that the
age estimates based on the MSTO are reliable, and hence, that the GC
age estimates are free of systematic errors.


\section{Summary}\label{sec6}
Globular clusters are the oldest objects in the universe which can be
dated.  Absolute GC ages based on the luminosity of the MSTO are the
most reliable \cite{renzini} and lead to the tightest constraints on
the age of the universe.  MSTO ages for GCs determined by a number of
different researchers agree well with each, and have not appreciable
changed for a number of years \cite{ageref}.  A number of independent
tests (summarized in \S \ref{sec5}) of the stellar evolution models
suggest that current stellar models are a good representation of
actual stellar evolution, and hence, that there are no unknown
systematic errors in the GC age estimates.  A detailed study of the
known uncertainties has lead to the conclusion that oldest GCs are
$14.6\pm 1.7\,$Gyr old, with a one-sided, 95\% confidence limit that
these clusters are older than $12.1\,$Gyr \cite{cdkk}.  To this age,
one must add some estimate for the time it took GCs to form after the
Big Bang.  Estimates for this formation time vary from 0.1 -- 2 Gyr.
To be conservative, the lower value is chosen.  Thus, the universe
must be older than $12.2\,$Gyr.  This mimimum value of $t_o$ requires
that $H_o < 72\,{\rm km/s/Mpc}$ if $\Omega = 0.1,\, \Lambda = 0$, or
$H_o < 54\,{\rm km/s/Mpc}$ for a flat, dominated universe ($\Omega =
1.0,\, \Lambda = 0$).

\section*{Acknowledgments}
Parts of this review are based on work done with P.\ Demarque, P.\
Kernan, and L.\ Krauss (Monte Carlo of GC ages), and Sarbani Basu and
J.\ Christensen-Dalsgaard (helioseismology).  I am grateful to Ata
Sarajedini for furnishing Figure 1.

\end{document}